\documentclass[iop,apj,tighten]{emulateapj}
\usepackage{amsmath,amstext}
\usepackage{apjfonts}
\usepackage[breaklinks,colorlinks,citecolor=blue,linkcolor=magenta]{hyperref} 
\usepackage[all]{hypcap} 
\usepackage{float}
\usepackage{footmisc}
\usepackage{multirow}
\usepackage{lineno}

\def \nustar {{\em NuSTAR}}
\def \xmm {{\em XMM-Newton}}
\def \suzaku {{\em Suzaku}}
\def \chandra {{\em Chandra}}
\def \rxte {{\em RXTE}}

\def \swift {{\em Swift}}
\def \inte {{\em INTEGRAL}}
\def \igr {{IGR~J17091--3624}}
\def \grs {{GRS~1915+105}}

\def \xillver{{\tt xillver}}
\def \relxill{{\tt relxill}}

\newcommand{\Msun}      {\mbox{$M_{\mathord\odot}$}}

\shorttitle{\igr\ Hard State}
\shortauthors{Xu et al.}

\begin{document}

\title{Spectral and Timing Properties of IGR J17091--3624 in the Rising Hard State During its 2016 Outburst}

\author{Yanjun Xu\altaffilmark{1}}
\author{Javier A. Garc\'ia\altaffilmark{1,2,3}}
\author{Felix F\"urst\altaffilmark{4}}
\author{Fiona A. Harrison\altaffilmark{1}}
\author{Dominic J. Walton\altaffilmark{5}}
\author{John A. Tomsick\altaffilmark{6}}
\author{Matteo Bachetti\altaffilmark{7}}
\author{Ashley L. King\altaffilmark{8}}
\author{Kristin K. Madsen\altaffilmark{1}}
\author{Jon M. Miller\altaffilmark{9}}
\author{Victoria Grinberg\altaffilmark{10}}

\altaffiltext{1}{Cahill Center for Astronomy and Astrophysics, California Institute of Technology, Pasadena, CA 91125, USA}
\altaffiltext{2}{Remeis Observatory \& ECAP, Universit\"at Erlangen-N\"urnberg, Sternwartstr.~7, 96049 Bamberg, Germany}
\altaffiltext{3}{Harvard-Smithsonian Center for Astrophysics, 60 Garden St., Cambridge, MA 02138 USA}
\altaffiltext{4}{European Space Astronomy Centre (ESA/ESAC), Operations Department, Villanueva de la Ca\~nada (Madrid), Spain}
\altaffiltext{5}{Institute of Astronomy, University of Cambridge, Madingley Road, Cambridge CB3 0HA, UK}
\altaffiltext{6}{Space Sciences Laboratory, 7 Gauss Way, University of California, Berkeley, CA 94720-7450, USA}
\altaffiltext{7}{INAF/Osservatorio Astronomico di Cagliari, via della Scienza 5, I-09047 Selargius (CA), Italy}
\altaffiltext{8}{KIPAC, Stanford University, 452 Lomita Mall, Stanford, CA 94305, USA}
\altaffiltext{9}{Department of Astronomy, University of Michigan, 1085 South University Avenue, Ann Arbor, MI 48109, USA}
\altaffiltext{10}{ESA/ESTEC, Keplerlaan 1, 2201 AZ Noordwijk, The Netherlands}

\begin{abstract}
We present a spectral and timing study of the \nustar\ and \swift\ observations of the black hole candidate IGR J17091--3624 in the hard state during its outburst in 2016. Disk reflection is detected in each of the \nustar\ spectra taken in three epochs. Fitting with relativistic reflection models reveals that the accretion disk is truncated during all epochs with $R_{\rm in}>10~r_{\rm g}$, with the data favoring a low disk inclination of $\sim 30^{\circ}-40^{\circ}$. The steepening of the continuum spectra between epochs is accompanied by a decrease in the high energy cut-off: the electron temperature $kT_{\rm e}$ drops from $\sim 64$~keV to $\sim 26$~keV, changing systematically with the source flux. We detect type-C QPOs in the power spectra with frequency varying between 0.131~Hz and 0.327~Hz. In addition, a secondary peak is found in the power spectra centered at about 2.3 times the QPO frequency during all three epochs. The nature of this secondary frequency is uncertain, however a non-harmonic origin is favored. We investigate the evolution of the timing and spectral properties during the rising phase of the outburst and discuss their physical implications.

\end{abstract}

\maketitle

\section{INTRODUCTION}
Most Galactic black hole binaries detected to date are transient X-ray sources that exhibit recurrent bright outbursts. During the onset of the month-to-year long outbursts, the sources transition from the low/hard state to the high/soft state through relatively short-lived intermediate states. In the hard state, the sources typically have hard spectra with a power-law index $\Gamma \sim$1.7 and the thermal disk component is cool and faint (usually not detected above 2~keV). In the frequency domain, type-C quasi-periodic oscillations (QPOs) ranging in frequency from $\sim 0.1-10$ Hz have been frequently detected, and they are considered as potential probes of the strong-gravity dominated flow dynamics \cite[see][for relevant reviews]{bhb_rev06, bhb_rev12}.

The standard paradigm of accretion disks in black hole binaries is that the inner disk radius extends to the innermost stable circular orbit (ISCO) in the soft state, whereas it is truncated at a larger distance during the hard state. There are two methods widely used to measure the inner disk radius from X-ray spectral modeling: determining the disk emission area from its thermal disk blackbody component \citep[e.g.,][]{zhang97, mcc14}, or estimating the relativistic distortion effects on the disk reflection features \citep[e.g.,][]{fab89, rey14}. 

While it is generally well supported by observations that the disk radius extends to the ISCO at high accretion rates in the soft state \citep[e.g,][]{steiner10, tom14, wal16}, the disk truncation interpretation of the hard state spectrum is still highly debated \citep[e.g.,][]{kol14,furst15,gar15}. Cool disks are common in the brightest phases of the hard state; modeling thermal continuum emission in this state is difficult, but some continua point to small inner disk radii \citep[e.g.,][]{Miller06,ryk07,rey13}. Evidence that disks may remain at the ISCO in bright phases of the hard state also comes from reflection spectroscopy \citep[e.g,][]{miller15}. In addition, it has been noticed that the measurements of truncated disks could be biased by photon pile-up distortions in some cases \citep[e.g.,][]{miller10}, with the best evidence of disk truncation (a narrow Fe K line) only observed in GX~339-4 by \suzaku\ at the luminosity approximately an order of magnitude below the brightest parts of the hard state \citep{tom09}. With the high sensitivity of \nustar\ for reflection spectroscopy and its triggered read-out \citep{nustar}, new observations in the hard state can help to better understand the disk evolution of black hole binaries.

\igr\ is a transient Galactic black hole candidate discovered with \inte\ in 2003 \citep{kuu03}. During its brightest outburst in 2011 \citep{burst11}, \igr\ was regularly monitored with \swift\ and \rxte. The source revealed a rich variety of variability behavior, with both low- and high-frequency QPOs detected \citep{alt11b,rod11,alt12}. After following the canonical evolution track for a black hole candidate at the beginning of the outburst, the source entered the soft state and was found to display a number of peculiar variability patterns. Most notably, the so-called "heartbeat" or $\rho$ state was detected, which had only been seen previously in the bright Galactic binary \grs\ \citep{alt11}. In addition, several new variability states were recently discovered in \igr\ that had never been reported in \grs\ \citep{pah12,court17}.  

Despite its similar timing properties to \grs, \igr\ is more than 10--50 times fainter. Assuming that the exceptional variability patterns are a result of emitting close to the Eddington limit, \igr\ either hosts an extremely small black hole $<3$~\Msun\ (for a distance of less than 17~kpc), or is very distant \citep{alt11}. However, both scenarios are in tension with other independent observations: for example, based on the empirical relations between the photon index and QPO frequency combined with constraints from spectral modeling, the black hole mass of \igr\ is estimated to be in the range of 8.7\Msun\ to 15.6 \Msun\ \citep{iyer15}; meanwhile, the source distance inferred from simultaneous multi-wavelength observations is between $\sim$11 to $\sim$17~kpc \citep{rod11}. Assumptions about other properties of the system, such as a low black hole spin or a high disk inclination have been invoked to help reconcile this discrepancy \citep{cap12, rao12}. So far, with no direct measurement of the basic properties of \igr\ (the distance, the inclination and the black hole mass and spin all remain unknown), it is difficult to distinguish among the different scenarios proposed and investigate the physical origin of its peculiar behavior.

In February 2016, \igr\ started to show renewed activity \citep{burst16}, and subsequent Swift monitoring confirmed the source rose through the hard state before transitioning to the soft state at a comparable flux level with its outburst in 2011. In this paper, we perform spectral and timing analyses of three \nustar\ and simultaneous \swift\ observations taken during the rising phase of the hard state. In Section \ref{sec:data}, we describe the observations and the data reduction. Section \ref{sec:ana} provides the details of our spectral fitting and timing analyses. We present a discussion in Section~\ref{sec:dis} and summarize our results in Section~\ref{sec:con}.

\section{OBSERVATIONS AND DATA REDUCTION}
\label{sec:data}
\begin{figure}
\centering
\includegraphics[width=0.49\textwidth]{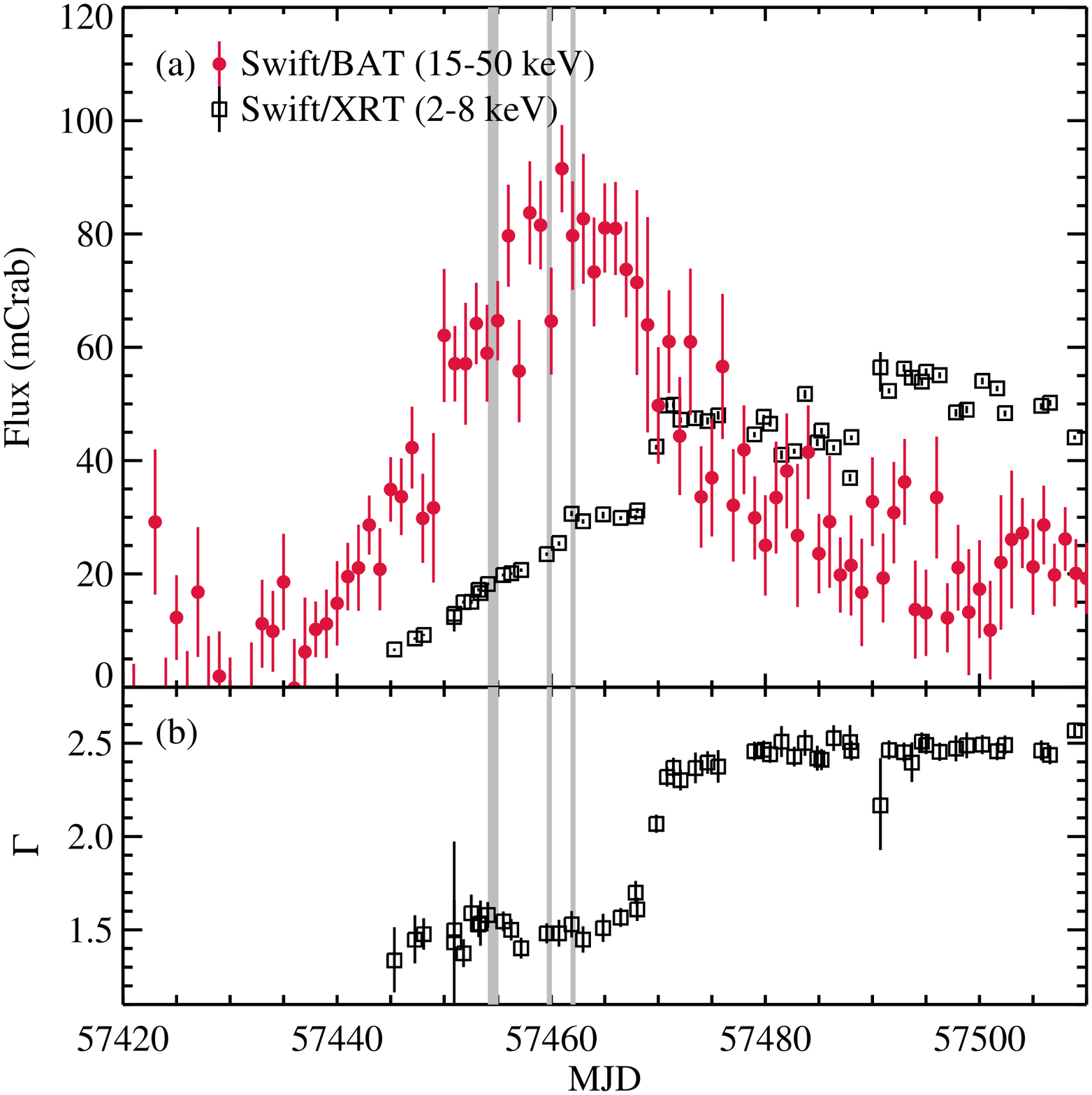}
\caption{(a) \swift\ monitoring of the 2016 outburst of \igr. The 15--50~keV BAT light curve is from the Swift/BAT hard X-Ray monitor, which is rescaled to the unit mCrab based on the Crab rates in the instrument band \citep{krimm13b}. The XRT light curve is converted to mCrab from the energy flux in 2--8 keV. (b) Photon index derived by fitting the \swift/XRT data with an absorbed power-law model. The gray shaded areas mark the three \nustar\ observations, which caught the source in the hard state before it started to enter the soft state around MJD 57470.
\label{fig:fig1}}
\end{figure}

\igr\ was observed by \nustar\ \citep{nustar} three times in the hard state during the rising phase of its 2016 outburst, on March 7, 12 and 14, about one week before the source started to enter the soft state (see Table~\ref{tab:tab1} for the list of observations considered in this work). \swift/XRT monitored the entire outburst of \igr\ with frequent 1--2~ks snapshots from February 26 to October 4. Figure~\ref{fig:fig1} shows the evolution of the soft and hard X-ray band flux during the outburst measured by the Swift X-ray Telescope (XRT) \citep{swiftxrt} and Burst Alert Telescope (BAT) \citep{swiftbat}. The \swift/XRT data were taken in the Windowed Timing mode to avoid photon pile-up. There are simultaneous \swift/XRT data for all the three \nustar\ observations, providing coverage of the soft X-ray band (see Table~\ref{tab:tab1}).

\capstartfalse
\begin{deluxetable}{clllll}[]
\tablewidth{\columnwidth}
\tablecolumns{7}
\tabletypesize{\scriptsize}
\tablecaption{\nustar\ and Simultaneous \swift\ Observations \label{tab:tab1}}
\tablehead{
\colhead{Epoch} & 
\colhead{Instrument}  & 
\colhead{ObsID}  &  
\colhead{Start Time} & 
\colhead{E. T.} &
\colhead{C. R.} \\
 & 
 & 
 &  
\colhead{(UTC)} & 
\colhead{(ks)} &
\colhead{(s$^{-1}$)}
} 

\startdata
 1       &  \nustar      & 80001041002  & 2016-03-07 02:01  & 43.3  & 16.5 \\
         &  \swift/XRT   & 00031921099  & 2016-03-07 01:49  & 1.46  &  7.3\\
\noalign{\smallskip}                               
 2       &  \nustar      & 80202014002  & 2016-03-12 14:11  & 20.2  & 21.4\\
         &  \swift/XRT   & 00031921104  & 2016-03-12 13:53  & 1.95  & 8.9 \\
\noalign{\smallskip}                               
 3       &  \nustar      & 80202014004  & 2016-03-14 19:26  & 20.7  & 23.6 \\
         &  \swift/XRT   & 00031921106  & 2016-03-14 21:39  & 1.03  &  10.4
\enddata
\tablecomments{
The exposure time and count rates of \nustar\ quoted here are the values from FPMA.
}
\end{deluxetable}

\subsection{NuSTAR}
We processed the \nustar\ data using v.1.6.0 of the NuSTARDAS pipeline with \nustar\ CALDB v20170120. After the standard data cleaning and filtering, the total exposure time is 43.3~ks, 20.2~ks and 20.7~ks for the three observations. For each observation, the source spectra and light curves were extracted at the position of \igr\ within the radius of 100$\arcsec$ from the two \nustar\ focal plane modules (FPMA and FPMB). Corresponding background spectra were extracted from source-free areas on the same chip using polygonal regions. In the first \nustar\ observation (ObsID 80001041002), the source region was contaminated by the stray light from a nearby bright source GX 349+2. In this case, we also placed the background region in the part of the field illuminated by the stray light. We note the influence from the stray light is minimal, as the background only contributes $\sim$2\% to the total count rate. The \nustar\ spectra were grouped to have a signal-to-noise (S/N) ratio of at least 20 per bin after background subtraction.

For the \nustar\ timing analysis, we first applied barycenter correction to the event files to transfer the photon arrival times to the barycenter of the solar system using JPL Planetary Ephemeris DE-200. We then generated cross-power density spectra (CPDS) using MaLTPyNT \citep{timingcode} from the cleaned event files. MaLTPyNT, developed for \nustar\ timing analysis, properly treats orbital gaps and uses the CPDS as a proxy for the power density spectrum. The CPDS uses the signals from two independent focal plane
modules, FPMA and FPMB, and the real part of the CPDS provides a measure of the signal that is in phase between the two modules. Thus, this method helps to remove the spurious distortion to the power-density spectrum introduced by the instrumental dead time \citep[for details, see][]{nustartiming}. We generated the CPDS with the binning time of 2$^{-8}$~s and in 512~s long intervals. For all the power spectra, we used the root-mean-square (rms) normalization, and geometrically rebinned the power spectra with a factor of 1.03 to increase the S/N of the frequency bins and produce nearly equally spaced bins in the logarithmic scale.

\subsection{Swift}
The \swift/XRT data were processed using {\tt xrtpipeline} v.0.13.2 with XRT CALDB v20160609 to produce clean event files. We extracted the source spectra with {\tt xselect} from a circular region with a radius of 20 pixels centered on the position of \igr. The background was extracted from an annulus area with the inner and outer radii of 80 pixels and 120 pixels, respectively. The data were filtered for grade 0 events only and we used {\tt swxwt0s6\_20131212v015.rmf} for the response matrix.  Ancillary response files were generated with {\tt xrtmkarf} incorporating exposure maps to correct for the bad columns. Finally, we rebinned the data to have a S/N greater than 5 for each energy bin.

\section{ANALYSIS}
\label{sec:ana}
\subsection{Spectral Modeling}
We model the simultaneous \nustar\ and \swift/XRT spectra in XSPEC v12.9.0n \citep{xspec} using $\chi^{2}$ statistics. For spectral fitting, we adopt the cross-sections from \cite{crosssec} and abundances from \cite{wil00}. All parameter uncertainties are reported at the 90\% confidence level for one parameter of interest. Cross-normalization constants are allowed to vary freely for \nustar\ FPMB and \swift/XRT and assumed to be unity for FPMA. The values are within ∼5\% of unity for Epoch 1 and Epoch 2, as expected from \cite{madsen15}. In Epoch 3, the \swift\ normalization is $\sim$13\% higher than \nustar. We note it is probably caused by the exposure correction uncertainty when the source lies close to the CCD bad columns, which is sensitive to the exact source position and can be affected by the spacecraft pointing stability\footnote{http://www.swift.ac.uk/analysis/xrt/digest\_sci.php}.

\begin{figure}
\centering
\includegraphics[width=0.47\textwidth]{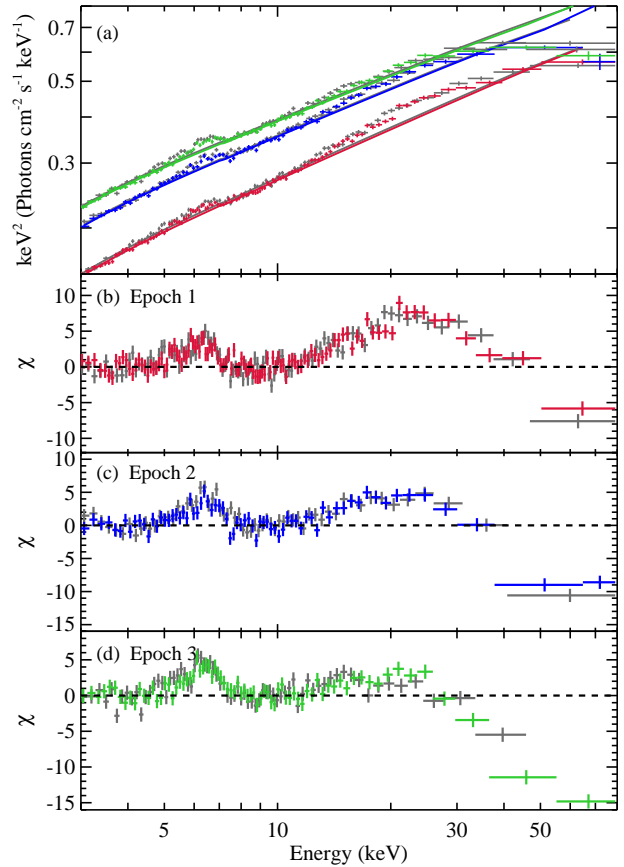}
\caption{(a) Unfolded \igr\ spectra from the three \nustar\ observations. (b)--(d) Residuals in $\chi$ from the simple absorbed power-law model. FPMA spectra taken in Epoch 1, 2, and 3 are marked in red, blue and green, respectively. For comparison, FPMB spectra are all in gray. The data are rebinned for display clarity. 
\label{fig:fig2}}
\end{figure}

All three \nustar\ observations detect \igr\ across the entire instrument bandpass. To highlight the reflection features, we fit the \nustar\ spectra with a simple absorbed power-law model, {\tt TBabs*powerlaw} in XSPEC, using only data in the 3--5~keV and 8--12~keV energy range. As shown in Figure~\ref{fig:fig2}, the residuals display asymmetric broad iron lines around 4--7~keV and show evidence for changes in the cut-off energy as the source luminosity increases. From this simple model, we also detect a slight steepening in the power-law slope: $\Gamma$ increases from $\sim$1.55 to $\sim$1.60 from Epoch 1 to Epoch 3. No parameter uncertainties are calculated here as the fits are too poor to determine meaningful parameter uncertainties.\\

\subsubsection{Joint Fits with Disk Reflection Models}
\label{3.1.1}
Given the disk reflection features observed (see Figure~\ref{fig:fig2}), we model the \nustar\ and \swift\ spectra simultaneously with the self-consistent reflection model \relxill\ v0.5b \citep{relxilla, relxillb}. We fit the \nustar\ and \swift\ spectra in the bands 3--79~keV and 1--10~keV, respectively. We ignore the \swift\ data below 1~keV to avoid possible low-energy residuals known to be present in the XRT Windowed Timing mode data of some absorbed sources\footnote{http://www.swift.ac.uk/analysis/xrt/digest\_cal.php}. We note that due to the relative weakness of the disk reflection component, as displayed in Figure~\ref{fig:fig2}a, reflection features are not detected by \swift. Although measurements of the reflection spectra are dominated by the \nustar\ data, extending the spectral fitting to the soft energy band helps to constrain the absorption column density and the general continuum slope.

\begin{figure}
\centering
\includegraphics[width=0.49\textwidth]{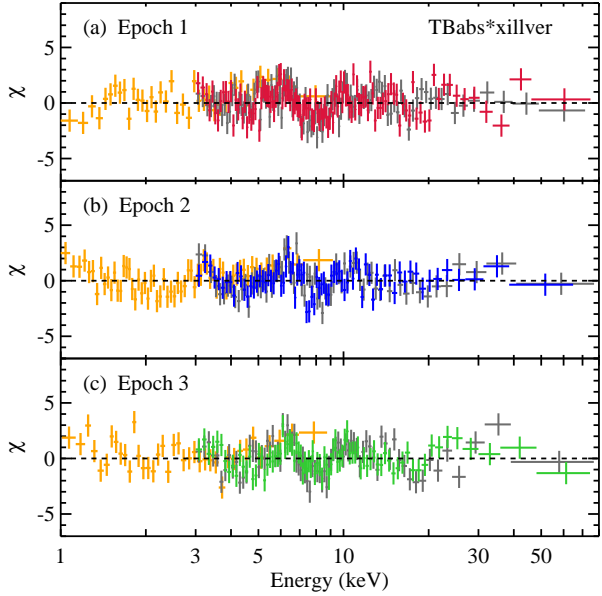}
\caption{Residuals of the unblurred reflection model. The same color scheme is used as in Figure~\ref{fig:fig2} for the \nustar\ data. \swift/XRT spectra are all plotted in yellow. The plots are in units of $\chi$ so that \swift\ residuals are of comparable magnitude with \nustar.
\label{fig:fig3}}
\end{figure}

We first fit the data with the unblurred version of the reflection model \xillver\ \citep{garcia10} multiplied by a model that
accounts for neutral Galactic absorption: {\tt TBabs*xillver}. To maximize the parameter constraints, we jointly fit the spectra taken at the three epochs and link the values for the absorption column density $N_{\rm H}$, the iron abundance $A_{\rm Fe}$ and the disk inclination $i$, which are normally expected to remain constant. Other relevant parameters (the ionization parameter $\xi$, photon index $\Gamma$, high energy cut-off $E_{\rm cut}$ and reflection fraction $R_{\rm ref}$) are allowed to vary between epochs. The best-fit result yields the reduced-chi-square $\chi^2_{\nu} = \chi^2/\nu = 3851.1/3629=1.061$, where $\nu$ is the number of degrees of freedom. The model leaves obvious residuals in the Fe K band (see Figure~\ref{fig:fig3}), suggesting that relativistic blurring is required to provide an adequate fit for the data.

\begin{figure}
\centering
\includegraphics[width=0.49\textwidth]{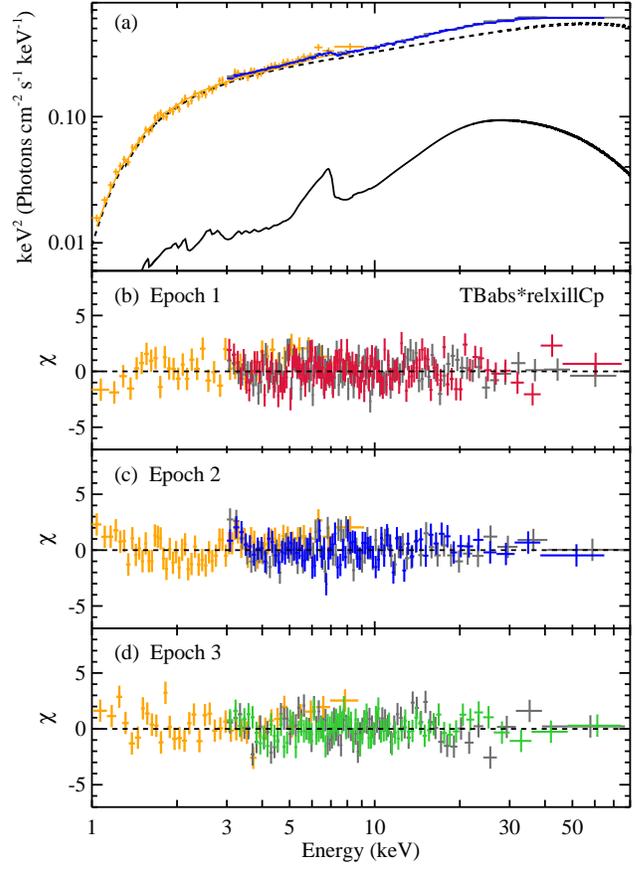}
\caption{(a) Unfolded \nustar\ and \swift\ spectra with Model la. For clarity, only the Epoch 2 data are displayed. Black solid and dashed lines indicate the relative contributions from the disk reflection and the coronal continuum in Model 1a. The normalization of the plotted model is the value corresponding to FPMA. (b)--(d) Residuals of Model 1a at the three epochs. The same color scheme is used as in Figure~\ref{fig:fig3}. 
\label{fig:fig4}}
\end{figure}

\capstartfalse \begin{deluxetable*}{clllllllllllll}[]
\tablewidth{\textwidth}
\tablecolumns{13}
\tabletypesize{\scriptsize}
\tablecaption{Best-fit Parameters of the Disk Reflection Models \label{tab:tab2}}
\label{fig1}
\tablehead{
\colhead{Epoch} 
&\colhead{$N_{\rm H}$} 
&\colhead{$q_{\rm in}$} 
&\colhead{$q_{\rm out}$} 
&\colhead{$h$} 
&\colhead{$i$} 
&\colhead{$R_{\rm in}$} 
&\colhead{a} 
&\colhead{$\Gamma$} 
&\colhead{log~${\xi}$} 
&\colhead{$A_{\rm Fe}$}
&\colhead{$k_{\rm Te}$} 
&\colhead{$R_{\rm ref}$}\\
&\colhead{($\times 10^{22}$ cm$^2$)} 
& 
&
&\colhead{($r_{\rm g}$)} 
&\colhead{($^\circ$)} 
&\colhead{($r_{\rm g}$)} 
& 
& 
&\colhead{(log[erg~cm~s$^{-1}$])} 
&
&\colhead{(keV)} 
&}

\startdata
\noalign{\smallskip}
\multicolumn{13}{c}{Model 1a: \textsc{TBabs*relxillCp}} \\
\noalign{\smallskip}
\hline
\noalign{\smallskip}
 1  & $(1.59\pm{0.03})^l$  & 3$^{*}$  & 3$^{*}$ & \nodata   & $(37^{+8}_{-5})^l$  & $20_{-8}^{+33}$  &0.998$^{*}$ &  $1.674_{-0.009}^{+0.008}$   & $1.95_{-0.18}^{+0.41}$  & $(0.74^{+0.10}_{-0.15})^l$  & $64_{-15}^{+46}$   & $0.26\pm{0.03}$          \\  
\noalign{\smallskip}                                                                                  
 2  & \nodata              & 3$^{*}$  & 3$^{*}$ & \nodata   & \nodata             & $17_{-7}^{+14}$  &0.998$^{*}$ &  $1.680_{-0.010}^{+0.012}$   & $2.73_{-0.21}^{+0.08}$  &\nodata                       & $32_{-4}^{+6}$    & $0.19_{-0.04}^{+0.06}$   \\
\noalign{\smallskip}                                                                                
 3  & \nodata              & 3$^{*}$  & 3$^{*}$ & \nodata   & \nodata             & $40_{-22}^{+47}$ &0.998$^{*}$ &  $1.698_{-0.008}^{+0.009}$   & $2.75\pm{0.09}$         &\nodata                       & $26_{-2}^{+3}$    & $0.15_{-0.03}^{+0.04}$   \\
\noalign{\smallskip}
\hline
\noalign{\smallskip}
\multicolumn{7}{r}{$\chi^2$/$\nu$}       & \multicolumn{6}{l}{3627.0/3626(1.000)} \\
\noalign{\smallskip}
\hline
\noalign{\smallskip}
\multicolumn{13}{c}{Model 1b: \textsc{TBabs*relxillCp} ($i$ allowed to vary between epochs)} \\
\noalign{\smallskip}
\hline
\noalign{\smallskip}
 1  & $(1.56\pm{0.03})^l$  & 3$^{*}$  & 3$^{*}$  & \nodata  & $34^{+8}_{-6}$   & $17_{-6}^{+24}$  &0.998$^{*}$ &  $1.672\pm{0.009}$           & $2.07_{-0.26}^{+0.34}$  & $(0.74^{+0.16}_{-0.14})^l$  & $69_{-7}^{+11}$  & $0.24_{-0.03}^{+0.04}$           \\  
\noalign{\smallskip}                                                                                                                                                                                                  
 2  & \nodata              & 3$^{*}$  & 3$^{*}$  & \nodata  & $42^{+11}_{-7}$  & $21_{-10}^{+20}$ &0.998$^{*}$ &  $1.683_{-0.010}^{+0.012}$   & $2.72_{-0.30}^{+0.07}$  &\nodata                      & $33_{-4}^{+6}$   & $0.21_{-0.05}^{+0.07}$    \\
\noalign{\smallskip}                                                                                                                                                                                                 
 3  & \nodata              & 3$^{*}$  & 3$^{*}$  & \nodata  & $29^{+21}_{-8}$  & $21_{-9}^{+77}$  &0.998$^{*}$ &  $1.696_{-0.009}^{+0.010}$   & $2.79_{-0.08}^{+0.16}$  &\nodata                      & $26_{-2}^{+3}$   & $0.13_{-0.03}^{+0.04}$    \\
\noalign{\smallskip}
\hline
\noalign{\smallskip}
\multicolumn{7}{r}{$\chi^2$/$\nu$}       & \multicolumn{6}{l}{3623.7/3624(1.000)} \\
\noalign{\smallskip}
\hline
\noalign{\smallskip}
\multicolumn{13}{c}{Model 2a: \textsc{TBabs*relxilllpCp}} \\
\noalign{\smallskip}
\hline
\noalign{\smallskip}
 1  & $(1.59\pm{0.03})^l$ & \nodata      & \nodata   & $9.4^{+29.3}_{-4.6}$    & $(36^{+5}_{-2})^l$  & $19_{-8}^{+21}$  &0.998$^{*}$ &  $1.673\pm{0.008}$          & $2.04_{-0.29}^{+0.31}$   & $(0.73\pm{0.13})^l$     & $67_{-17}^{+39}$ & $0.51$ \\  
\noalign{\smallskip}                                                                                                                                                                                                      
 2  & \nodata             & \nodata      & \nodata   & $5.7^{+6.5}_{-3.0}$     & \nodata             & $18_{-7}^{+8}$   &0.998$^{*}$ &  $1.680_{-0.009}^{+0.010}$   & $2.74_{-0.20}^{+0.07}$   &\nodata                 & $33_{-4}^{+6}$   & $0.40$  \\
\noalign{\smallskip}                                                                                                                                                                                                      
 3  & \nodata             & \nodata      & \nodata   & $11.9^{+20.2}_{-7.8}$   & \nodata             & $39_{-20}^{+21}$ &0.998$^{*}$ &  $1.698\pm{0.008}$           & $2.75_{-0.08}^{+0.09}$  &\nodata                 & $26_{-2}^{+3}$    & $0.31$ \\
\noalign{\smallskip}
\hline
\noalign{\smallskip}
\multicolumn{7}{r}{$\chi^2$/$\nu$}       & \multicolumn{6}{l}{3626.9/3626(1.000)}  \\
\noalign{\smallskip}
\hline
\noalign{\smallskip}
\multicolumn{13}{c}{Model 2b: \textsc{TBabs*relxilllpCp} (R$_{in}$ linked for all epochs)} \\
\noalign{\smallskip}
\hline
\noalign{\smallskip}
 1  & $(1.59\pm{0.03})^l$ & \nodata   & \nodata   & $10.3^{+3.3}_{-2.5}$    & $(35\pm{4})^l$      & $(21_{-6}^{+11})^l$  &0.998$^{*}$ &  $1.673_{-0.008}^{+0.006}$   & $2.04_{-0.29}^{+0.30}$  & $(0.72^{+0.12}_{-0.13})^l$  & $67_{-17}^{+35}$ & $0.50$  \\  
\noalign{\smallskip}                                                                                                      
 2  & \nodata             & \nodata   & \nodata   & $6.6^{+7.3}_{-3.0}$     & \nodata             & \nodata              &0.998$^{*}$ &  $1.678\pm{0.009}$          & $2.74_{-0.19}^{+0.08}$  &\nodata                      & $33_{-4}^{+5}$   & $0.36$   \\
\noalign{\smallskip}                              
 3  & \nodata             & \nodata   & \nodata   & $5.5^{+7.1}_{-2.4}$     & \nodata             & \nodata              &0.998$^{*}$ &  $1.700_{-0.008}^{+0.009}$   & $2.79_{-0.06}^{+0.10}$  &\nodata                      & $26_{-2}^{+3}$   & $0.31$  \\
\noalign{\smallskip}
\hline
\noalign{\smallskip}
\multicolumn{7}{r}{$\chi^2$/$\nu$}       & \multicolumn{6}{l}{3629.3/3628(1.000)} 
\enddata
\tablecomments{
Frozen parameters are marked with asterisks. Superscripts $l$ indicate the parameters whose values are linked at the three epochs. For all models, the outer disk radius R$_{\rm out}$ is fixed at 400 r$_{\rm g}$ and the black hole spin $a$ is fixed at 0.998. As discussed in the text, altering the spin parameter $a$ causes no significant change to the fitting result.
}
\end{deluxetable*}

\subsubsection{Power-law Emissivity Index}
In order to account for the relativistic effects and to better describe the changes in the high energy rollover as displayed in Figure~\ref{fig:fig1}, we fit the spectra with the relativistic reflection model {\tt relxillCp} (Model 1a). It uses a thermally Comptonized input continuum {\tt nthcomp} \citep{zdz96,zyc99} and parameterizes the high energy cut-off by the coronal electron temperature $kT_{\rm e}$. The disk emissivity profile in {\tt relxillCp} is described by a broken power law with three parameters: the inner and the outer emissivity indices $q_{\rm in, out}$ and the break radius $R_{\rm br}$. We assume a canonical emissivity profile $\epsilon(r)\varpropto r^{-3}$ by fixing both $q_{\rm in}$ and $q_{\rm out}$ at 3, which is expected for the outer parts of a standard Shakura--Sunyaev disk \citep{dau13}. If $q_{\rm in}$ and $R_{\rm br}$ are allowed to vary individually for each epoch, the fit would not be improved and the data cannot constrain the extra parameters. We fix the outer disk radius at 400~$R_{\rm g}$ and allow the inner disk radius $R_{\rm in}$ to vary to test the disk truncation hypothesis. There is no well-constrained black hole spin measurement for \igr; previous estimates range from a low or negative spin \citep{rao12} to a high spin \citep{reis12}. We first simply fix the dimensionless spin parameter $a$ ($a \equiv cJ/GM^{2}$) at its maximum value of 0.998. As demonstrated below, the choice of spin does not significantly affect our results.

\begin{figure*}
\centering
\includegraphics[width=1.00\textwidth]{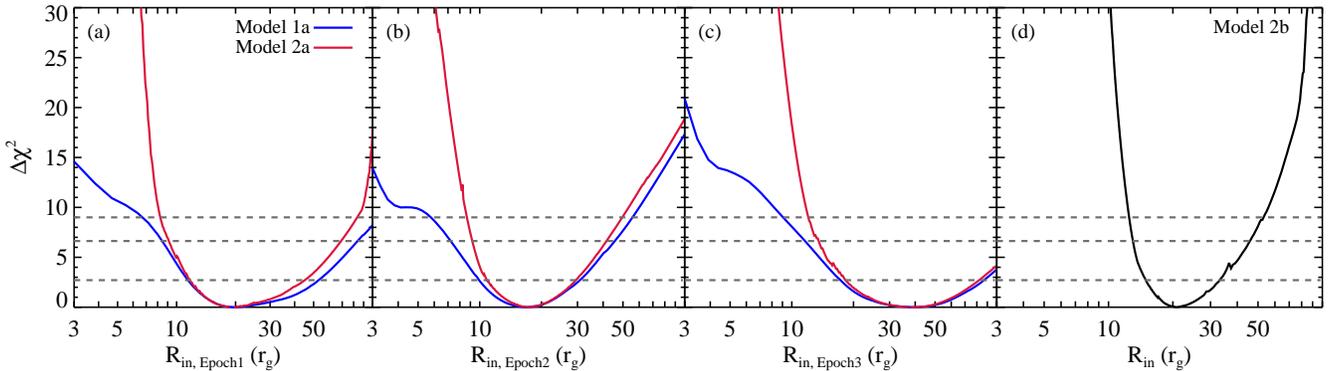}
\caption{$\Delta\chi^2$ contours of the inner disk radius $R_{\rm in}$ obtained from Model 1a, 2a and 2b. The dashed lines indicate the 90\%, 99\% and 3$\sigma$ confidence levels for one parameter of interest.
\label{fig:fig5}}
\end{figure*}

The broad-band X-ray spectra can be well fitted by Model 1a with no obvious residuals (Figure~\ref{fig:fig4}). The reduced-chi-square is very close to unity ($\chi^2_{\nu}=3627.0/3626=1.000$), indicating statistically a very good fit. Using the relativistic reflection model improves the fit significantly by $\Delta \chi^2=224$ compared to the unblurred reflection model. The only excess noticeable is around 3.0--3.5~keV in the \nustar\ spectra at Epoch 2 (Figure~\ref{fig:fig4}c), which is possibly related to small calibration uncertainties near the edge of the instrument bandpass and is not statistically important. We include this part in the spectral modeling as it helps to constrain the overall power-law slope. The soft disk component is either too cool to contribute much above 1~keV or intrinsically too faint to be detected by \swift/XRT, and a disk blackbody component is not required by our spectral modeling. From the best-fit parameters, we confirm the relatively high obscuration in \igr, with the equivalent hydrogen column density $N_{\rm H} = (1.58\pm0.03) \times 10^{22}$~cm$^{-2}$, consistent with previous measurements \citep[e.g.,][]{rod11,cap12}. After including the reflection component in the spectral fitting, we measure a slightly steeper power-law index ($\Gamma\simeq1.67-1.70$) compared to the simple absorbed power-law model. We confirm the existence of the high energy cut-off in the spectra, with the relatively low coronal temperature\footnote{Approximately, the exponential rollover energy $E_{\rm c}$ of the thermally Comptonized continuum is related to the the electron temperature $kT_{\rm e}$ as $E_{\rm c}=2-3~kT_{\rm e}$. } $kT_{\rm e}=64_{-15}^{+46}$~keV, $32_{-4}^{+6}$~keV, $26_{-2}^{+3}$~keV for Epoch 1, 2 and 3, respectively. 

As for the key reflection parameters: the ionization parameter $\xi$\footnote{\label{footnote}The ionization parameter $\xi=4\pi F_{\rm x}/n$, where $F_{\rm x}$ is the ionizing flux and $n$ is the gas density.} increases with source flux from Epoch 1 to Epoch 3; an abundance of $A_{\rm Fe} = 0.78^{+0.10}_{-0.13}$ (in units of solar value) is derived for the accretion disk  and the disk is estimated to be viewed at a low inclination angle $i = 37^{\circ +3}_{-4}$; the measured inner disk radius is $R_{\rm in,Epoch1}=20_{-8}^{+33}$~$r_{\rm g}$, $R_{\rm in,Epoch2}=17_{-7}^{+14}$~$r_{\rm g}$ and $R_{\rm in,Epoch3}=40_{-22}^{+47}$~$r_{\rm g}$ (see Table~\ref{tab:tab2} for a list of the best-fit parameters).

The accretion disk is considered truncated if the inner edge of the disk is located outside the ISCO. The radius of the ISCO is a function of the black hole spin, and its value decreases monotonically from 9~$r_{\rm g}$ ($r_{\rm g} \equiv GM/C^2$ is the gravitational radius) for an extreme retrograde spinning black hole to 1.235~$r_{\rm g}$ for a black hole with a maximum positive spin. The lower limits of the inner disk radius from Model 1a are all in excess of 10 $r_{\rm g}$ at the 90\% confidence level and radii smaller than 6~$r_{\rm g}$ can be ruled out at 3$\sigma$ (see Figure~\ref{fig:fig5}), which points to the truncated disk scenario.

The reflection fraction $R_{\rm ref}$ is defined to be the ratio of the coronal intensity illuminating the disk to that reaching the observer \citep{dauser16}. The low reflection fraction measured ($R_{\rm ref}\sim 0.15-0.26$) suggests the photon reprocessing is free from the strong light-bending effects in the vicinity of the black hole, also consistent with a truncated accretion disk. To investigate the effect of the black hole spin on the fitting results, we fix $a$ at -0.998, 0 and 0.998 respectively, and obtain basically identical fits for all other parameters. Therefore, we note that the reflection modeling is simply not sensitive to the spin parameter, which is to be expected when the disk is truncated at this level. By fixing $R_{\rm in}$ at the ISCO and in turn fitting for the black hole spin $a$ would yield a worse fit by $\Delta\chi^2 \simeq 20$ with $a$ pegged at -0.998. This indicates that the fitting is attempting to find a larger inner disk radius than the maximum value allowed for the ISCO, which is also evidence for a truncated disk.

In Model 1a, we assume the iron abundance, inclination and absorption column density remain constant through the outburst, which can be an over-simplification in some cases. For instance, a strongly warped disk could cause the apparent disk inclination to vary with radius. Evidence for disk warping has been found in e.g., Cyg X--1, for which the inner disk and the orbital plane inclination are found to disagree by $\sim 10^{\circ}-15^{\circ}$ \citep{tom14, wal16}. We explore this possibility by fitting independently the disk inclination parameter $i$ at the three epochs (Model 1b). This only improves the fitting marginally with two extra free parameters $\Delta\chi^2/\Delta\nu=-4.3/-2$. The best-fit inner disk radius at Epoch 3 is a bit lower, about the same value at Epoch 2 (although still consistent with Model 1a results within errors). This is more reasonable, as an obvious increase of the inner radius returned by Model 1a is not expected given the short time interval between the later two observations. However, since there is no clear indication for a change in the inner disk radius between epochs, disk warping is not explicitly required to describe the data. 

\subsubsection{Lamppost Geometry}
The characteristic reflection spectra in black hole binaries are believed to be produced by reprocessing of the hard X-ray continuum generated by a centrally located corona. In principle, information about the geometry of the corona can be extracted from the disk reflection spectrum. In the previous section we measured a change in the electron temperature $kT_{\rm e}$, implying an evolving corona. In an attempt to further explore the possible changes in the accretion geometry, we assume a lamppost geometry for the corona by using the {\tt relxilllpCp} reflection model. Instead of assigning a specific emissivity profile, {\tt relxilllpCp} characterizes the reflection spectrum assuming a point source located in the rotational axis at a height $h$. The lamppost model {\tt relxilllpCp} can self-consistently calculate the reflection fraction $R_{\rm ref}$ given a certain combination of spin, inner disk radius and coronal height, which places an extra constraint on the fitting \citep[see][for a discussion]{relxilla}. We first fix the spin parameter $a$ at the maximum, and allow both the inner disk radius and the lamppost height to vary freely between epochs (Model 2a). The lamppost model results in a good fit to the data with $\chi^2_{\nu}=3626.9/3626=1.000$. The best-fit parameters are in general consistent with the results from the {\tt relxillCp} model, and we find a coronal height of $\sim10~r_{\rm g}$ (see Table~\ref{tab:tab2}). As the disk is truncated and given the data quality, the modeling is not sensitive to the high emissivity close to the black hole that would result from a low lamppost height. The inner disk radius and the lamppost position cannot be simultaneously constrained to test a plausible change in the parameters.  

Since the values of the inner disk radius at the three epochs are all consistent within errors, we make an attempt to achieve a better constraint on $h$ by tying together $R_{\rm in}$ at the three epochs in a joint fit (Model 2b). This yields an equally acceptable fit with $\chi^2_{\nu}=3629.3/3628=1.000$. The coronal height $h$ is measured to be $h_{\rm Epoch 1}=10.3^{+3.3}_{-2.5}~r_{\rm g}$, $h_{\rm Epoch 2}=6.6^{+7.3}_{-3.0}~r_{\rm g}$ and $h_{\rm Epoch 3}=5.5^{+7.1}_{-2.4}~r_{\rm g}$, with $R_{\rm in}$ staying at $21^{+11}_{-6}~r_{\rm g}$. We note the constraint on $h$ here is most likely driven by the decrease of the reflection fraction from Epoch 1 to Epoch 3. As the spin and the inner disk radius are both assumed to remain constant, in the lamppost model the different reflection fractions can only be accounted for by the change in the lamppost position. In this case, the reflection fraction returned by our model fits correlates positively with the the coronal height. This trend is opposite to that expected if the disk extends down to the ISCO. When the accretion disk is truncated, more photons are lost with a lower coronal height, because light-bending in the strong gravitational field focuses more X-rays to the innermost regions that will not be reflected by the truncated disk. This trend of the reflection fraction, also observed in the fits with Model 1a, is yet another indication for the moderate truncation of the
accretion disk.

\begin{figure}
\centering
\includegraphics[width=0.49\textwidth]{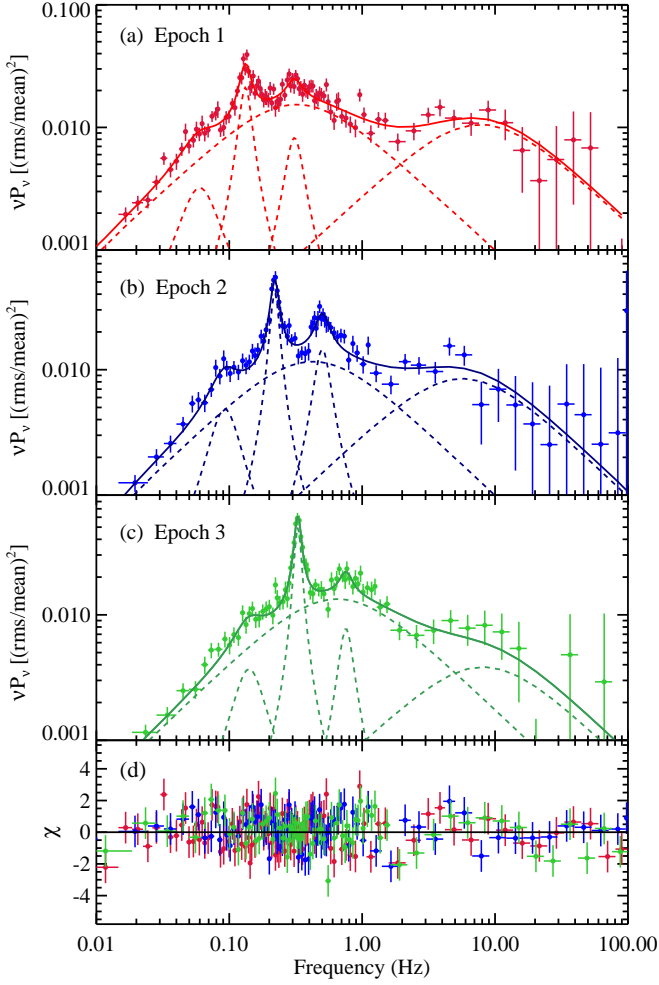}
\caption{(a)--(c) \nustar\ CPDS and best-fit model in the $\nu P_{\nu}$ representation. The CPDS are calculated using data from the full energy band (3--79~keV). Same color scheme is used as in the previous figures: data taken in Epoch 1, 2, 3 are plotted in red, blue and green, respectively. A type-C QPO variable in frequency along with a secondary peak around 2.3 times the QPO frequency are detected. The dashed lines indicate the best-fit Lorentzian profiles. (d) Residuals of the best-fit model that includes five Lorentzians. The data are rebinned for display clarity.
\label{fig:fig6}}
\end{figure}

\subsection{Timing Analysis} 

\begin{figure}
\centering
\includegraphics[width=0.49\textwidth]{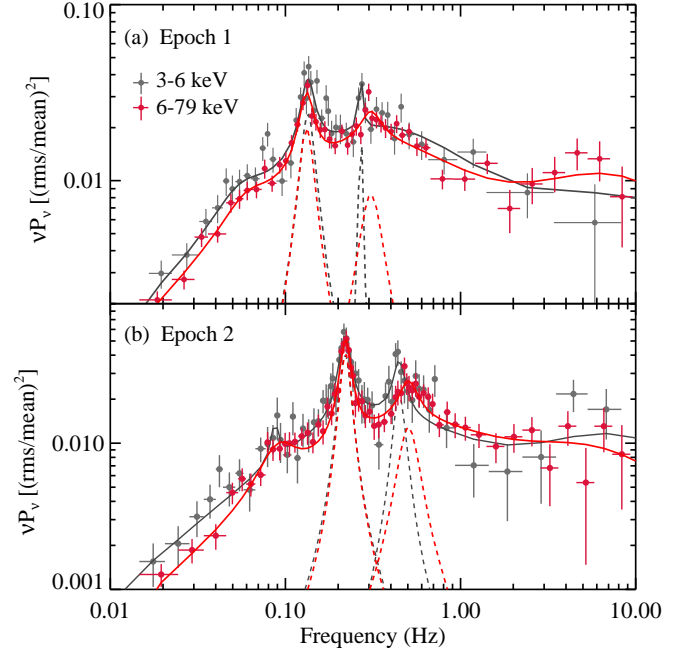}
\caption{CPDS generated in two energy intervals: 3--6~keV (gray) and 6--79~keV (red). Panel (a) and (b) are the data taken from the the first and the second \nustar\ observations. The dotted lines indicate the best-fit Lorentizian profiles for the type-C QPO and the secondary frequency. 
\label{fig:fig7}}
\end{figure}

Low frequency QPOs observed in black hole binaries are classified into different types, namely type-A, type-B and type-C, based on their characteristics in the power spectra and are related to different spectral states (e.g., see \citealt{cas04, cas05}). In the hard state, type-C QPOs are most commonly found, which are variable and highly coherent peaks on top of strong flat-topped noise. A type-C QPO is evident in the CPDS generated in the full energy band (3--79~keV) from our \nustar\ observations of \igr\  (see Figure~\ref{fig:fig6}).

We model the \nustar\ CPDS of the three epochs in XSPEC jointly with a unit response file. We use a multi-Lorentizan model, which is commonly used to fit the power spectra of black hole binaries \citep[e.g.,][]{nowak00,pott03}. The model consists of two zero-centered Lorentzians for the underlying broad component, plus one Lorentzian for the fundamental QPO and two for possible sub- and super- harmonics. This provides an adequate fit for the power spectra, with $\chi^2_{\nu} = 829.9/798= 1.04$, leaving no obvious excess (see Figure~\ref{fig:fig6}). The Lorentzian used to fit for the possible sub-harmonic is less significant compared to other components, and the centroid frequency and the line width cannot be very well constrained. Therefore, we focus our discussion on the two more significant peaks in the CPDS. The type-C QPO frequency $\nu_{0}$ is well measured to be $0.131\pm{0.002}$~Hz, $0.219\pm{0.002}$~Hz and $0.327\pm{0.002}$~Hz, with the rms amplitude of $6.1\pm0.6$\%, $7.3\pm0.5$\% and $7.3\pm0.4$\% for Epoch 1 to 3, respectively. No significant deviation in the QPO frequency is observed during any of the three individual \nustar\ observations. The relatively low type-C QPO frequencies are consistent with the values found during the onset of the 2011 outburst of \igr\ \citep{rod11, iyer15}. However, by adopting the disk reflection model, we measure a slightly higher photon index of $\Delta \Gamma \sim 0.2$ from the energy spectra when about the same QPO frequencies are detected, compared to \cite{iyer15} who used different spectral models. 

As displayed in Figure~\ref{fig:fig6}, there is a secondary peak at higher frequencies than the type-C QPO in all three datasets. Such feature is commonly associated with the first QPO harmonic, which occurs at twice the fundamental QPO frequency. However, this secondary frequency $\nu_{1}$ is measured to be centered at $ 0.31\pm0.01$~Hz, $0.50\pm0.01$~Hz and $0.76^{+0.05}_{-0.04}$~Hz at Epoch 1, 2 and 3, respectively. $\nu_{1}$ is about 2.3 times the type-C QPO frequency, making them unlikely to be harmonically related. Although some small residuals can been seen in the fit of the power-spectra (e.g., around 0.3 Hz in Figure~\ref{fig:fig6}(b)), they are probably because the underlying continuum cannot be perfected described by two Lorenzians and would not affect the measurement of the secondary frequency. When modeling the CPDS, the Lorentzian width of the secondary peak cannot always be constrained within a reasonable value. In order to obtain a reasonable fit, we set an upper limit of 0.2~Hz for the Lorentzian FWHM. This feature is detected at $\gtrsim 4\sigma$ confidence level in all epochs, and is most significant at Epoch 2 with the detection at $7.6 \sigma$\footnote{All significances here are estimated with the F-test, based on the improvement in $\chi^2$ by adding an Lorentizan component for the secondary peak in the fit model. For low significance cases, we test the significances further with Monte Carlo simulations using the $\tt simftest$ script in $\tt xspec$. The results agree with those calculated by the simple F-test.}. Although its origin is unclear, we note this is not caused by instrumental effects, as its characteristic frequency is increasing between epochs.

\capstartfalse
\begin{deluxetable}{ccllll}[]
\tablewidth{\columnwidth}
\tablecolumns{6}
\tabletypesize{\scriptsize}
\tablecaption{Fitting Results of the \nustar\ CPDS in Different Energy Bands \label{tab:tab3}}
\tablehead{
\colhead{Epoch} & 
\colhead{Energy~(keV)}  & 
\colhead{$\nu_0$~(Hz)} &   
\colhead{$\nu_1$~(Hz)} &
\colhead{$P(\nu_{1})$}
} 

\startdata
 1       &  3.0--79.0      & $0.131\pm{0.002}$           & $0.31\pm0.01$ & $5.3\sigma$ \\
\noalign{\smallskip}  
         &  3.0--6.0       & $0.134^{+0.004}_{-0.003}$   & $0.270^{+0.006}_{-0.004}$ & $2.0\sigma~(2.4\sigma)$\\
\noalign{\smallskip}  
         &  6.0--79.0      & $0.131\pm{0.002}$           & $0.30^{+0.02}_{-0.01}$  & $3.9\sigma$\\
\noalign{\smallskip}  
\hline
\noalign{\smallskip}
 2       &  3.0--79.0      & $0.219\pm{0.002}$          & $0.50\pm0.01$  & $7.6\sigma$\\
\noalign{\smallskip}  
         &  3.0--6.0       & $0.218\pm0.004$            & $0.45^{+0.03}_{-0.02}$  & $3.6\sigma$\\
\noalign{\smallskip}  
		 &  6.0--79.0      & $0.220\pm0.002$            & $0.50\pm0.01$  & $ 5.1\sigma$\\
\noalign{\smallskip}                      
\hline   
\noalign{\smallskip}
 3       &  3.0--79.0      & $0.327\pm{0.002}$          & $0.76^{+0.05}_{-0.04}$  & $4.0\sigma$\\
\noalign{\smallskip}  
         &  3.0--6.0       & $0.327\pm{0.004}$          & \nodata  & $<95\%$~(79\%)\\
\noalign{\smallskip}  
		 &  6.0--79.0      & $0.327\pm{0.003}$          & $0.70^{+0.05}_{-0.04}$ & $2.7\sigma~(2.6\sigma)$
\enddata
\tablecomments{
$\nu_{0}$ and $\nu_{1}$ denote the type-C QPO and the secondary frequency. $P(\nu_{1})$ is the significance of the secondary frequency determined by the F-test. For comparison, significances calculated by simulations (1000 trials) are listed in parentheses.
}
\end{deluxetable}

We also investigate the energy dependence of the CPDS. In order to allow for enough statistics in each energy band, we only calculate the CPDS in two energy intervals: 3--6~keV (low) and 6--79~keV (high). The type-C QPO frequency basically remains constant in the two energy bands. The only noteworthy difference is evidence for an apparent shift of the secondary frequency, which is present at Epoch 1 and 2 (see Figure~\ref{fig:fig7}). In the high energy band, this secondary frequency is detected at $ 3.9\sigma$ and $5.1\sigma$ in Epoch 1 and Epoch 2, respectively; while it becomes considerably weaker at lower energies, the confidence levels are $2.0\sigma$ in Epoch 1 and $3.6\sigma$ in Epoch 2. The fitting finds different values for the secondary frequency $\nu_{\rm 1}$ in the two energy bands: at lower energies we measure $\nu_{\rm 1, Epoch1}=0.270^{+0.006}_{-0.004}$~Hz, $\nu_{\rm 1, Epoch2}=0.45^{+0.03}_{-0.02}$~Hz; whereas $\nu_{\rm 1}$ derived from the high energy band agrees with the full-band results (see Table~\ref{tab:tab3} for a list of frequencies and significances). We note $\nu_1$ in the 3--6~keV band is roughly two times the type-C QPO frequency, consistent with that of a conventional first harmonic. If the secondary frequency $\nu_{\rm 1}$ is required to be the same in the two energy bands, the overall fit would worsen by $\Delta \chi^2_{\rm Epoch1} \simeq 8$, $\Delta \chi^2_{\rm Epoch2} \simeq 5$. In the Epoch 3 data, the secondary peak is the weakest and cannot be detected above the 95\% confidence level in the low energy band. The absence of the same pattern in Epoch 3 could imply it is transient, or it just lacks the statistics to be detected as the power spectrum becomes noisy close to the high frequency end. In addition, we calculate the lag-frequency spectra between the two energy bands. The time lags are consistent with zero around the QPO frequencies.

\begin{figure}
\centering
\includegraphics[width=0.50\textwidth]{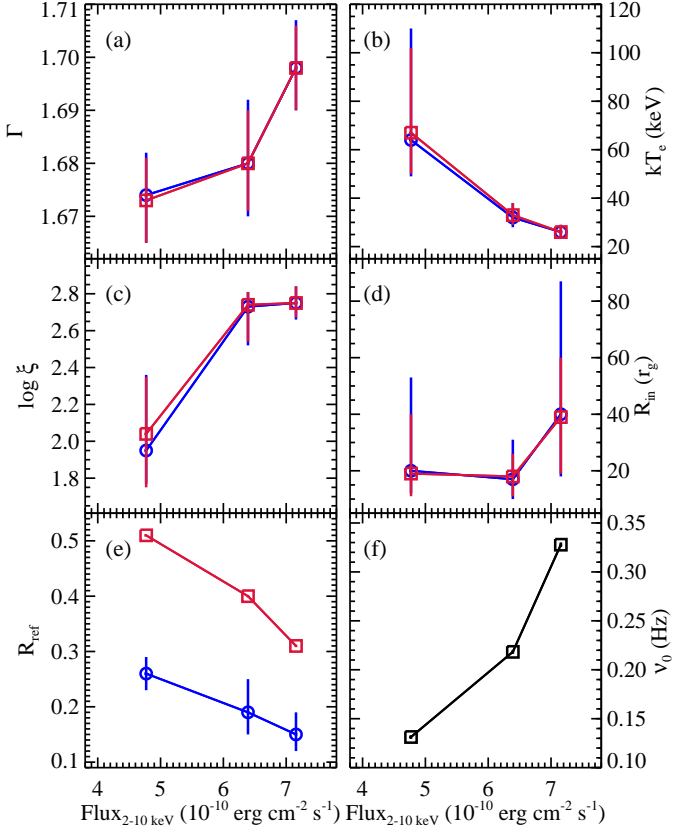}
\caption{Relation of spectral and timing properties with the source flux. Blue circles the are best-fit parameters from Model 1a, red squares are the results from Model 2a that assumes a lamppost geometry. There are no error bars for the reflection fraction $R_{\rm ref}$ from Model 2a in panel (e), as the values are calculated self-consistently by the model from a combination of $a$, $R_{\rm in}$ and $h$.
\label{fig:fig8}}
\end{figure}

\section{DISCUSSION}
We have presented a multi-epoch X-ray analysis of the galactic black hole candidate \igr\ in the rising hard state using data taken by \nustar\ and \swift. The data is free from pile-up issues at the count rate level during the observations. The broadband X-ray spectra display broadly similar shapes, revealing a blurred reflection component on top of a cut-off power-law continuum, with no significant detection of the soft disk emission. We model the multi-epoch energy spectra jointly with four different relativistic reflection model setups, assuming either a constant emissivity index or a steep emissivity profile from a lamppost geometry. The models all provide adequate fits for the data and yield consistent results (see Table~\ref{tab:tab2}). During the spectral modeling, we allow key parameters governing the inner disk properties to vary between epochs. Timing analysis of the \nustar\ data reveals a type-C QPO in the power spectra that increases in frequency from Epoch 1 to Epoch 3. We also detect a secondary frequency at about 2.3 times the type-C QPO frequency, inconsistent with the conventional frequency expected from QPO super-harmonics, possibly an independent second frequency of the system.

\label{sec:dis}
\subsection{Properties of the Inner Accretion Disk}
Our reflection modeling finds the accretion disk to be truncated. The lower limits of the inner disk radius are all in excess of $10~r_{\rm g}$, which is larger than the maximum ISCO radius corresponding to a retrograde spin. We have also demonstrated that the choice of black hole spin has negligible effects on the measurement of disk truncation in our case. The best-fit values for the inner disk radius are around $17-40~r_{\rm g}$. Considering the parameter uncertainties, the truncation radius is at the level of a few tens of gravitational radii. Due to controversies over the source distance and the black hole mass, the Eddington ratio of the source during our observations is uncertain. It can be estimated if we assume \igr\ emits at the Eddington limit during its "hearbeat" state. From \cite{alt11}, the 2--50 keV source flux was $\sim4\times10^{-9}$~erg~cm$^{-2}$~s$^{-1}$ when \grs\ like variabilities were detected. For a crude estimation, taking a bolometric correction factor of 3, we find an Eddington fraction $\lambda=L_{\rm 1-100~keV}/L_{\rm Edd}\sim 20-30\%$ at the time of the \nustar\ observations (the source flux only increases by $\sim$50$\%$ from Epoch 1 to 3 as shown in Figure~\ref{fig:fig8}). Therefore, our results are currently consistent with the disk still being truncated at a considerable level during the bright phase of the outburst ($L/L_{\rm Edd} \gg 1\%$), although we stress that the exact value for the Eddington ratio is highly uncertain.

We obtain an abundance of $A_{\rm Fe}\simeq0.7$ for the accretion disk from the spectral fitting. Super-solar abundances of $A_{\rm Fe}\sim 3-5$ have been frequently required in the reflection modeling of Galactic black hole binaries: e.g., Cyg X--1 \citep{furst15,parker15,furst16,wal16}, GX 339--4 \citep{tom14,gar15} and GRS 1739--278 \citep{miller15}. The lower elemental abundance for \igr\ could result from the weak Fe K line/edge relative to the Compton hump.

A face-on geometry is favored for \igr\ with a low inclination of $i \simeq 30^{\circ}-40^{\circ}$. A high inclination ($> 65^{\circ}$) can be excluded by all models at $>90\%$ confidence level (see Table~\ref{tab:tab2}), and in the best constrained case (Model 2b), this can be ruled out at $> 5 \sigma$. A low inclination is in contrast with several previous estimations. \cite{cap12} proposed \igr\ could be a highly inclined system, as the faintness of the source could be ascribed to the spectral deformation effects due to a high inclination angle. \cite{rao12} performed a phase-resolved spectral fitting of the heartbeat oscillations of \igr\ observed with \rxte\ and \xmm\ with a multi-temperature disk blackbody and a power-law component, and their results favor a high inclination of $\sim$70$^{\circ}$. In addition, high velocity outflows have been observed in \igr\ by \chandra\ \citep{king12}, that are usually expected from nearly edge-on systems. Suggestions of a high inclination for \igr\ also come from its general similarities with \grs, that has been measured to have an inclination angle of $\sim 65^{\circ}-75^{\circ}$ \citep[e.g.,][]{mid06, miller13}.

A warped accretion disk could be one possible explanation for this discrepancy. All previous arguments for the high inclination are based on the soft state properties of \igr, while our measurements indicating a low inclination are from the hard state. From the standard diagram for black hole binaries, the inner disk radius is believed to move inward towards the ISCO during the transition from the hard state to the soft state, the inner edge of the accretion disk can appear to be viewed at different angles if the disk is strongly warped: the inner part is aligned with the black hole spin while the outer part is tilted, aligned with the binary orbital axis. This misalignment is believed to be driven by the precessional torque from the Lense-Thirring effect \citep{bard75}. Location of the transition of orientations can reach a steady state and is at $\sim 10-20$~$r_{\rm g}$ as shown by recent simulations \citep{neal15}, which is relevant to the scale of truncation radius discussed here. Therefore, the viewing angle $30^{\circ}-40^{\circ}$ we measured might be the orbital inclination or it might be somewhere in between the orbital and inner disk inclinations. Unfortunately, as we have not detected significant changes in the inner disk radius considering the error range, it is not possible to test this hypothesis via spectral fitting within the timespan of our observations.

\subsection{Origin of the Secondary Frequency}
Regarding the nature of the peak around $2.3$ times the fundamental frequency, the basic question is whether it is harmonically related to the type-C QPO frequency or not. Harmonics of low-frequency QPOs have been discovered from a number of Galactic black hole binaries, and have been used as an additional probe to investigate the physical origins of QPOs \citep[e.g.][]{axel14,ingr15,axel16}. While the fundamental QPO and its harmonics have been found to display different behaviors (e.g., the energy dependence of phase lags and amplitudes), the ratio of the fundamental and first harmonic frequency has been always found at 1:2 \citep[e.g.,][]{rod02,cas05,ingr15}. However, based on the measurements from the full-band CPDS of \igr, the frequency ratio of the secondary peak and the type-C QPO $\nu_{\rm 1}/\nu_0$ is $2.37\pm0.08$, $2.28\pm0.05$, $2.32\pm0.02$ for Epoch 1 to 3, respectively, making it inappropriate to associate the secondary frequency with the first QPO harmonic. 

In the low energy band (3-6 keV) CPDS, the secondary peak is found at the conventional value of two times the QPO frequency. We note that instead of being an apparent shift of the secondary frequency detected in the full band, it is possible that this is the real first harmonic. It is not unusual for the super-harmonic to be more prominent in the lower energy band while it is undetected at higher energies, as the harmonic energy spectrum has been found to be systematically softer than that of the fundamental QPO. This is likely a result of the inhomogeneity of the Comptonizing region generating the QPOs \citep{axel14,axel16}. Also, the harmonic features are usually transient, which could explain why it is not detected in the soft band CPDS at Epoch 3.  

The secondary frequency at about 2.3 times the type-C frequency is consistently detected in the 6-79 keV and full energy band, indicating that it is dominated by hard X-ray photons. One possibility is that higher energy photons are generated at a smaller distance from the black hole where the times scales are shorter, corresponding to a higher QPO frequency. Pairs of non-harmonically related QPOs have been simultaneously discovered during the 2005 outburst of GRO J1655--40, which were identified as a type-C QPO and a type-B QPO \citep{motta12}. However, the secondary frequency detected in \igr\ is unlikely a type-B QPO, since type-B QPOs are normally detected at higher frequencies around $\sim 5-6$~Hz and when the source is at a higher hardness ratio \citep[e.g.,][]{cas04,cas05}. Although GRO J1655--40 is not an exact analog to that discussed here, it indicates that there are multiple mechanisms generating the low-frequency QPOs in black hole binaries. Currently there is no consensus about the nature of low frequency QPOs. If the secondary frequency discussed here is indeed not harmonically related with the type-C QPO, its origin is rather uncertain but it should be an independent frequency that increases with the type-C QPO frequency.

\subsection{Evolution of the Spectral and Timing Properties}
The 2--10~keV flux of \igr\ increased by $\sim$50\% during the time span of our three \nustar\ observations. We observe several parameters from the spectral and timing analysis to change systematically with the source flux (see Figure~\ref{fig:fig8}). 

The evolution in the spectral fitting parameters is broadly similar to what has been found in the systematic study of GX 339-4 recently with \rxte\ \citep{gar15}. The ionization parameter increases with the source flux, which is naturally expected given the definition of the ionization parameter $\xi$\footref{footnote}. Regarding the two parameters governing the general curvature of the continuum, the photon index $\Gamma$ increases whereas the electron temperature $kT_{\rm e}$ decreases with rising source flux. The slope of the thermal Comptonized continuum is described by an asymptotic power-law with photon index $\Gamma$ in the {\tt nthcomp} model, which depends on the electron scattering optical depth $\tau_{\rm e}$ and the electron temperature $kT_{\rm e}$ \citep{lightman87}. Therefore, optical depth $\tau_{\rm e}$ can be estimated from $\Gamma$ and $kT_{\rm e}$\footnote{We use equation (1) in \cite{gar15} for the calculation.} and we find $\tau_{\rm e, Epoch1}\simeq1.95$, $\tau_{\rm e, Epoch2}\simeq3.13$, $\tau_{\rm e, Epoch3}\simeq3.51$. The measurements suggest the corona is cooling more efficiently by Compton up-scattering more photons as the source flux increases, which is also physically consistent with the increase in the electron opacity. 

\begin{figure}
\centering
\includegraphics[width=0.50\textwidth]{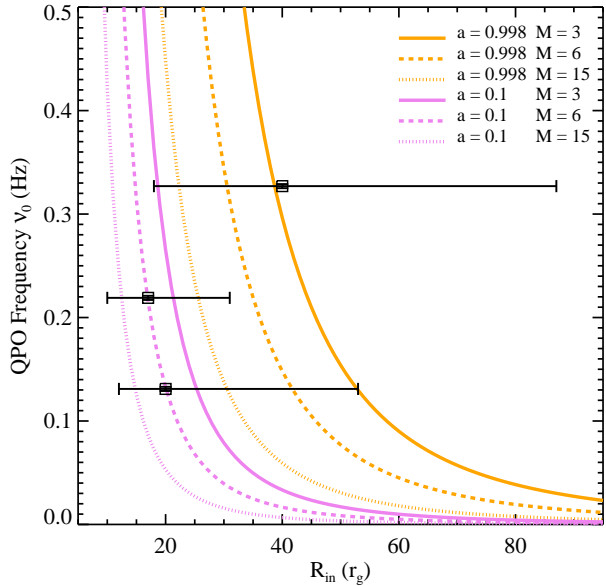}
\caption{Black square points mark the inner disk radius $R_{\rm in}$ and the type-C QPO frequency $\nu_0$ measured at the three epochs, the errors are clearly dominated by the measurement of disk truncation. For comparison, theoretical curves of the nodel precession frequency based on the Lense-Thirring precession model are plotted.
\label{fig:fig9}}
\end{figure}

The type-C QPO frequency is known to be well correlated with the general changes in the energy spectra, e.g., photon index, strength of the thermal disk component \citep{bhb_rev06}. One promising explanation for the low-frequency QPO is the Lense-Thirring precession of an inner hot flow of a truncated disk, which correlates the increase of the QPO frequency with the inward motion of the inner edge of the accretion disk \citep[e.g.,][]{ingram09,ingram11}. The model assigns the nodal precession frequency to the origin of type-C QPOs. In principle, it is possible to test the consistency of this QPO model using spectral-timing methods. We make a simple comparison of the QPO frequency $\nu_{\rm 0}$ and the inner disk radius $r_{\rm in}$ to the theoretical calculation of the nodal precession frequency $\nu_{\rm nod}$ from \cite{ingram14}. Since neither the black hole spin $a$ nor mass $m$ is known for the system, we calculate the theoretical curve for several combinations of these parameters (see Figure~\ref{fig:fig9}). However, the inner disk radius measured at the three epochs overlaps with each other within errors and no clear trend can be inferred from the three data points. It is clear from Figure~\ref{fig:fig9} that in order to detect possible tension between the spectral and timing results in the frame of the QPO model, e.g., whether the inner disk radius changes in step with the QPO frequency, the inner disk radius must be constrained to be better than $\sim$5~$r_{\rm g}$.

The anti-correlation of the reflection fraction with the source flux has also been observed and discussed in \cite{gar15}, although we note the reflection fraction is defined differently here. From the lamppost geometry, the decrease of the reflection fraction in \igr\ can be self-consistently explained by a reduction of the lamppost height assuming the inner disk radius remains constant (as the case with Model 2b), which can still be valid if the inner disk is moving inward as predicted by the QPO model. This only works if the corona has a compact size compared to the level of disk truncation. In this scenario, because there is no inner disk to reflect the emission, a lower source height means that a larger fraction of the photons that would have been reflected by an inner disk reaching the ISCO are now lost in the gap between the disk and the horizon, reducing the reflection fraction. We note that the decrease of the reflection fraction could also be caused by the beaming effect of an outflowing corona \citep{bel99, mal01}. However, it is difficult to explore this scenario with our limited S/N data, since additional parameters such as the corona velocity and geometric extent should be considered.

\section{SUMMARY AND CONCLUSION}
\label{sec:con}
We have undertaken a multi-epoch analysis of the \nustar\ and \swift\ observations of the black hole candidate \igr\ in the rising hard state during its 2016 outburst. Reflection features are detected in the \nustar\ spectra, enabling us to constrain the inner accretion properties from relativistic reflection modeling. Reflection models assuming a constant emissivity index or a lamppost geometry yield consistent results: the accretion disk is truncated at the level of a few tens of gravitational radii at all epochs, and is viewed at a low inclination angle of $\sim 30^{\circ}-40^{\circ}$. Our modeling also implies that the reflection spectrum is not sensitive to the specific disk emissivity profile when the disk is truncated. Several parameters from the spectral fitting (the photon index $\Gamma$, the coronal temperature $kT_{\rm e}$, the ionization parameter $\xi$ and the reflection fraction $R_{\rm ref}$) are observed to evolve systematically with the source flux, which is consistent with the standard picture of a black hole binary going into an outburst.

A type-C QPO is robustly detected in the \nustar\ data, with the frequency varying from 0.131~Hz to 0.327~Hz. The Lense-Thirring precession model predicts moderate truncation regardless of the black hole mass and spin. It also predicts a relatively small change in $R_{\rm in}$ with QPO frequency. Unfortunately, there is no clear evidence for a corresponding change in the disk truncation radius predicted by the model, as the inner disk radius cannot be well constrained from the reflection modeling, but the results are generally consistent with the model within errors. A secondary peak is detected at around 2.3 times the type-C frequency in the power spectra at all epochs, which we note is unlikely to be a QPO harmonic, but instead an independent frequency of the system. If true, this would add to the peculiarity of \igr\ in its variability behavior. Lacking enough statistics for more detailed timing analysis, the nature of this secondary frequency remains uncertain. 

\acknowledgments{
D.J.W. acknowledges support from an STFC Ernest Rutherford Fellowship. This work was supported under NASA contract No.~NNG08FD60C and made use of data from the \nustar\ mission, a project led by the California Institute of Technology, managed by the Jet Propulsion Laboratory, and funded by the National Aeronautics and Space Administration. We thank the \nustar\ Operations, Software, and Calibration teams for support with the execution and analysis of these observations. This research has made use of the \nustar\ Data Analysis Software (NuSTARDAS), jointly developed by the ASI Science Data Center (ASDC, Italy) and the California Institute of Technology (USA).}

\bibliographystyle{yahapj}
\bibliography{igr_hs.bib}
\end{document}